\documentclass[prb,aps,amsmath]{revtex4} 
\usepackage{amssymb}
\usepackage{latexsym}
\usepackage{graphicx}

\usepackage{epsfig}

\usepackage{dcolumn}
\usepackage{amsmath}
\usepackage{amsfonts}
\usepackage{bm}

\newcommand{\be}{\begin{equation}}
\newcommand{\ee}{\end{equation}}
\newcommand{\bea}{\begin{eqnarray}}
\newcommand{\eea}{\end{eqnarray}}

\newcommand{\p}{\partial}
\newcommand{\s}{\sigma}

\newcommand{\la}{\langle}
\newcommand{\ra}{\rangle}
\newcommand{\rd}{\mbox{d}}
\newcommand{\ri}{\mbox{i}}
\newcommand{\re}{\mbox{e}}

\begin{document}
\title{Comment on ``A continuous topological phase transition between 1D antiferromagnetic spin-1 boson superfluids'' by W. Ji and X.-G. Wen}
\author{ A. M. Tsvelik}
\affiliation{Division of Condensed Matter Physics and Materials Science, Brookhaven National Laboratory, Upton, NY 11973-5000, USA}
 \date{\today } 
 
 \begin{abstract} 
 This is a comment on the preprint arXiv:1809.07771v2 by W. Ji and X.-G. Wen.
 
\end{abstract}


\maketitle

In the preprint ``A continuous topological phase transition between 1D antiferromagnetic spin-1 boson superfluids'' W. Ji and X.-G. Wen take the approach based on the models two-leg fermionic ladders \cite{wen}. The spin liquids they refer to emerge as low energy limits of models of coupled fermionic chains (the so-called fermionic ladders). Away from half filling the U(1) symmetric sector of the theory containing excitations of total charge decouples from the rest. The remaining part of the Hilbert space which has U(1)$\times$SU(2)$\times$ Z$_2$ symmetry can exist in different phases depending on the interactions. The authors are interested in gapped phases and argue that there is one among them which is topologically nontrivial (similar to the Haldane phase of S=1 antiferromagnetic chain) and consider transitions into topologically trivial phases. 

 The models of fermionic ladders have been intensively studied during more than 20 years and the authors re-derive some of the results obtained before. In particular, they use the fact that once the total charge  sector is factored out the model can be recast in terms of six Majorana fermions. This approach has been used before \cite{fisher,fisher2,controzzi, tsvelik}, see also \cite{shelton} for the case of purely spin ladders. The Majorana fermions can be introduced using the bosonization and subsequent refermionization. Namely, one uses the bosonization formulae for the right-moving fermionic fields with spin projection $\s/2$ located on  chains $p = \pm 1$ (see also \cite{maldacena}): 
\bea
R_{p\s} = \frac{\kappa_{p\s}}{\sqrt{2\pi a_0}}\exp[\ri\sqrt{\pi}(\varphi_c + p\varphi_f + \s\varphi_s + p\s\varphi_{sf})], ~~ p = \pm 1; \s = \pm 1,
\eea
and likewise for the left-moving ones with $\bar\varphi$ bosonic fields satisfying the standard commutation relations. Then the Majorana fermions are defined as 
\bea
\xi_{1a} +\ri\xi_{2a} = \frac{\eta_{a}}{\sqrt{\pi a_0}}\re^{\ri\sqrt{4\pi}\varphi_a}, ~~ a = c,f,s,sf.
\eea
Here $\kappa_{p\s}, \eta_a$ are Klein factors. The Majoranas can be naturally gathered into groups;  the fermions $\xi_{1s},\xi_{2s}, \xi_{1sf}$ constitute the SU(2) triplet from SU$_2$(2) Kac-Moody algebra, $\xi_{2sf}$ is a singlet under this algebra and $\xi_{1f},\xi_{2f}$ transform under U(1) (O(2)). 

 As was demonstrated many times, including the aforementioned papers, the most general low energy theory for the two-leg fermionic ladder with the U(1) factored out is the one of 6 massless Majorana fermions with four fermion interactions :

\bea
&&  {\cal L} = \sum_{i=1}^6\left[\bar\xi_i(\p_\tau-iv\p_x)\bar\xi_i + \xi_i(\p_\tau+iv\p_x)\xi_i \right]   +g_{cf}(\bar\xi_{f1}\xi_{f1} +\bar\xi_{f2}\xi_{f2})^2\nonumber\\
&& - (\bar\xi_{f1}\xi_{f1} +\bar\xi_{f2}\xi_{f2})\Big(g_{cs,+}\sum_{b= s1,s2,sf1}\bar\xi_b\xi_b+ g_{cs,-}\bar\xi_{sf2}\xi_{sf2}\Big) \nonumber\\
&& -g_{s,+}\sum_{s1,s2,sf1; a>b}(\bar\xi_a\xi_a) (\bar\xi_b\xi_b) - g_{s,-}\sum_{a=s1,s2,sf1}(\bar\xi_a\xi_a) (\bar\xi_{sf2}\xi_{sf2}).
\label{eq:majoranaexp}
\eea
There is indeed the SU$_2$(2)  critical point where the three Majorana fermions $s1,s2, sf1$ are massless; it realized when the bare couplings satisfy $g_{cs,-} =0, g_{s,-} =0, g_{s,+} <0$. Small deviations of $g_{cs,-}, g_{s,-}$ from zero drive the system into different phases of the ladder, but the perturbing operators include for fermions and not two, as is stated in \cite{wen}. The  case of small deviations was analyzed in \cite{controzzi}. 

 The phase diagram of doped fermionic ladder has also been studied \cite{fisher,fisher2,controzzi}. It includes two phases with quasi long range superconducting order and two phases with quasi long range charge density wave order. In \cite{controzzi} the corresponding order parameters which are bilinear in terms of the original fermions were expressed in the language of the SU(2)$_2\times$SU(2)$_2$ Wess-Zumino model. In this language they are products of bosonic exponents from the U(1) sector and spin S=1/2 operators from the orbital and spin SU$_2$(2) sectors respectively. The problem of topological order has not been raised, but in the view of existence of local order parameter operators sensitive to external perturbations existence of topological phases appears unlikely.


\begin{thebibliography}{99}
\bibitem{wen} W. Ji and X.-G. Wen, ``A continuous topological phase transition between 1D antiferromagnetic spin-1 boson superfluids'', arXiv:1809.07771v2. 
\bibitem{fisher} L. Balents, M. P. A. Fisher, "Weak Coupling Phase Diagram of the Two-Chain Hubbard Model", Phys. Rev. B{\bf 53}, 12133 (1996).
\bibitem{fisher2} H.H. Lin, L. Balents and M. P. A. Fisher, "Exact SO(8) Symmetry in the Weakly-Interacting  Two-Leg Ladder", Phys. Rev. B{\bf 58}, 1794 (1998).
\bibitem{controzzi} D. Controzzi and A. M. Tsvelik, `Excitation spectrum of doped 2-leg ladders: a field theory analysis'', Phys. Rev. B{\bf 72}, 035110 (2005).
\bibitem{tsvelik} A. M. Tsvelik, ``A Field Theory for a Fermionic  Ladder with Generic Intrachain Interactions'', Phys. Rev. B {\bf 83}, 104405 (2011).
\bibitem{shelton} D. G. Shelton, A. A. Nersesyan and A. M. Tsvelik, " Antiferromagnetic Spin Ladders: Crossover between Spin S = 
1/2 and S = 1 Chains", Phys. Rev. B {\bf 53}, 8521  (1996). 
 \bibitem{maldacena} J. M. Maldacena and A. W. W. Ludwig, "Majorana Fermions, Exact Mapping between Impurity Fixed Points with four Bulk Fermionic Species, and Solution of the "Unitarity Puzzle", Nucl. Phys. B{\bf 506}, 565 (1997). 

\end{thebibliography}
\end{document}